\documentclass[english]{article}
\usepackage[T1]{fontenc}
\usepackage[latin9]{inputenc}
\usepackage{graphicx}
\usepackage{esint}

\makeatletter
\newcommand{\lyxaddress}[1]{
\par {\raggedright #1
\vspace{1.4em}
\noindent\par}
}

\makeatother

\usepackage{babel}
\begin{document}

\title{\textbf{On the equivalence between rotation and gravity: ``Gravitational''
and ``cosmological'' redshifts in the laboratory}}
\maketitle
\begin{center}
\textbf{\large{}Christian Corda}{\large\par}
\par\end{center}

\lyxaddress{\textbf{International Institute for Applicable Mathematics and Information
Sciences, B. M. Birla Science Centre, Adarshnagar, Hyderabad 500063
(India) and Istituto Livi, Via Antonio Marini, 9, 59100 Prato (Italy)}}

\lyxaddress{\textbf{E-mail: }\textbf{\emph{cordac.galilei@gmail.com.}}}
\begin{abstract}
The Mössbauer rotor effect recently gained a renewed interest due
to the discovery and explanation of an additional effect of clock
synchronization which has been missed for about 50 years, i.e. starting
from a famous book of Pauli, till some recent experimental analyses.
The theoretical explanation of such an additional effect is due to
some recent papers in both the general relativistic and the special
relativistic frameworks. In the first case (general relativistic framework)
the key point of the approach is the Einstein's equivalence principle
(EEP), which, in the words of the same Einstein, enables ``\emph{the
point of view to interpret the rotating system K' as at rest, and
the centrifugal field as a gravitational field}''. In this paper,
we analyse both the history of the Mössbauer rotor effect and its
interpretation from the point of view of Einstein's general theory
of relativity (GTR) by adding some new insight. In particular, it
will be shown that, if on one hand the ``traditional'' effect of
redshift has a strong analogy with the gravitational redshift, on
the other hand the additional effect of clock synchronization has
an intriguing analogy with the cosmological redshift. 

Finally, we show that a recent claim in the literature that the second
effect of clock synchronization does not exist is not correct.
\end{abstract}
\begin{quotation}
\end{quotation}

\section{Introduction}

Rotational motions always had and currently have an important role
in science in general and in physics in particular, starting from
Aristotelian metaphysics, passing through Newtonian physics and arriving
to the current relativistic paradigm. 

In the framework of the theory of relativity, the Sagnac effect, which
is due to the French physicist Georges Sagnac \cite{key-1,key-2},
has a historical importance as it shows the absolute character of
rotation by also enabling a long and interesting debate on the foundations
of the theory of relativity {[}1\textendash 10{]}, which involved
the same Einstein \cite{key-4,key-5}. 

In the context of the GTR, from the historical point of view it was
during his analysis of the rotating frame that Einstein had the intuition
to represent the gravitational field in terms of space-time curvature
\cite{key-11}. Einstein indeed wrote, verbatim \cite{key-11}:

``\emph{The following important argument also speaks in favor of
a more relativistic interpretation. The centrifugal force which acts
under given conditions of a body is determined precisely by the same
natural constant that also gives its action in a gravitational field.
In fact we have no means to distinguish a centrifugal field from a
gravitational field. We thus always measure as the weight of the body
on the surface of the earth the superposed action of both fields,
named above, and we cannot separate their actions. In this manner
the point of view to interpret the rotating system K' as at rest,
and the centrifugal field as a gravitational field, gains justification
by all means. This interpretation is reminiscent of the original (more
special) relativity where the pondermotively acting force, upon an
electrically charged mass which moves in a magnetic field, is the
action of the electric field which is found at the location of the
mass as seen by the reference system at rest with the moving mass.}''

This interpretation by Einstein of the rotating system in terms of
a gravitational field permitted various general relativistic analysis
of Mössbauer rotor experiments {[}12\textendash 17{]} and Sagnac experiments
\cite{key-18}. The key point of the above highlighted interpretation
by Einstein is the EEP which enables the equivalence between gravitation
and inertial forces {[}13\textendash 17{]}. The rotating reference
frame in a Mössbauer rotor experiment is included in the EEP {[}13\textendash 17{]}.
On the other hand, one must recall that, despite the importance that
gravitational physicists attribute to the EEP, one must be careful
when using it. The EEP has indeed \emph{local} validity and many limitations
when applied to extended regions of space. For example, the authors
of \cite{key-19} referring to the work \cite{key-20-2} pointed out
that, in general, there is no equivalence of inertial and gravitational
effects in a practical situation, in the context of solutions of Einstein's
equations. In addition, in \cite{key-21} it has been shown that it
is no longer possible to establish a frame with uniform acceleration
as equivalent to the homogeneous Newtonian gravitational field. In
the same ref. \cite{key-21} the authors mention the letter that Einstein
wrote to Max Plank, recognizing that the concept of ``uniform acceleration''
needs further clarification. 

In the framework of the GTR, the historical solutions of the Einstein
equations for the gravitational field of a rotating source are very
important. These are the Kerr \cite{key-22} and Kerr-Newman \cite{key-23}
solutions. On the other hand, Einstein, Lense and Thirring found an
interesting similarity between the gravitational field of a distribution
of mass and the electromagnetic field of a distribution of charge
\cite{key-24}. The Einstein-Thirring-Lense effect was the starting
point of the today popular framework of the gravitomagnetism see {[}25\textendash 31{]}
and references within. 

One must stress that relativistic rotation effects can be important
also in everyday life. It is indeed well known the use of the GTR
in Global Positioning Systems (GPS) \cite{key-32}. In that case,
the time difference between a frame co-rotating with the Earth geoid
and a fixed, locally inertial, Earth centered frame cannot be neglected
\cite{key-17,key-32}. In other words, GPS would not have the requested
accuracy if one neglects the relativistic effects due to the rotation
of the Earth \cite{key-17,key-32}. This issue will be also partially
clarified in this paper by confronting the GPS with the Mössbauer
rotor effect.

Rotation effects in astrophysics are important for what concerns the
famous Dark Matter problem. Such a problem originated in the 30's
of last century \cite{key-33}. If one observes the Doppler shift
of stars which move near the plane of the Milky Way and one calculates
the velocities, one finds a large amount of matter inside the Milky
Way preventing the stars from escaping out. Such a (supposed and unknown)
matter generates a very large gravitational force, that the luminous
mass in the Milky Way cannot explain. In order to achieve such a large
discrepancy, the sum of all the luminous components of the Milky Way
should be two or three times more massive. One can calculate the tangential
velocity of stars in orbits around the centre of the Milky Way like
a function of distance from the centre. The result will be that stars
which are far away from the centre of the Milky Way move with the
same velocity independent on their distance out from the centre. 

These puzzling issues generate a portion of the Dark Matter problem.
In fact, either luminous matter is not able to correctly describe
the radial profile of the Milky Way or, alternatively, Newton theory
of gravity cannot describe dynamics far from the centre of the Milky
Way. 

The dynamical description of various self-gravitating astrophysical
systems generates other issues of the Dark Matter problem. For example,
one can consider stellar clusters, external galaxies, clusters and
groups of galaxies. In such cases, the problem is similar. There is
more matter arising from dynamical analyses with respect to the total
luminous matter. 

Zwicky \cite{key-34} found that in the Coma cluster the luminous
mass is too little to generate the total gravitational field which
is needed to hold the cluster together. 

In any case, despite the Dark Matter problem is today much more complicated
than the sole issue of the rotation curves of the galaxies, one can
surely state that, historically, it arose from a rotational motion. 

In this paper an important rotational relativistic effect, the Mössbauer
rotor effect, will be analysed in a general relativistic framework.
The key point of the analysis will be the EEP, which, in the words
of the same Einstein, enables ``\emph{the point of view to interpret
the rotating system K' as at rest, and the centrifugal field as a
gravitational field}''.

In fact, the Mössbauer rotor effect recently gained a renewed interest
due to the discovery and explanation of an additional effect which
has been missed for about 50 years, i.e. starting from the book of
Pauli \cite{key-12}, till the more recent experimental analyses.
The theoretical explanation of such an additional effect is due to
the recent papers \cite{key-13,key-14,key-17} in the general relativistic
framework, and \cite{key-15} in the special relativistic framework.
In this paper, both the history of the Mössbauer rotor effect and
its interpretation from the point of view of Einstein's GTR are analysed
by adding some new insight. In particular, it will be shown that,
if on one hand the ``traditional'' effect of redshift has a strong
analogy with the gravitational redshift, on the other hand the additional
effect of clock synchronization has an intriguing analogy with the
cosmological redshift. 

In detail in this work we add the following new results to the discussion: 

i) In Section 3 of the paper we discuss a ``cosmological'' interpretation
of the time dilation due to relativistic contraction of the disc between
the co-rotating and inertial frames. In fact, in previous works \cite{key-13,key-14,key-17}
we merely derived an additional effect of clock synchronization, but
it was not completely clear that such an effect can be physically
interpreted in terms of a real blueshift. In this context, the analogy
with the cosmological redshift of the Universe strengthens previous
results \cite{key-13,key-14,key-17}, and could, in principle, generate
a further interest in the Mössbauer rotor effect.

ii) In Section 5 of the paper we show that a recent claim in the literature
that the second effect of clock synchronization does not exist \cite{key-35}
is not correct. This refutation of the claims in \cite{key-35}\emph{
}is needed for the sake of completeness.

\section{History of the Mössbauer rotor effect }

The Mössbauer effect takes its name by the discovery of the German
physicist R. Mössbauer in 1958 \cite{key-36}. Based on its importance
for various research fields in physics and chemistry, Mössbauer was
awarded the 1961 Nobel Prize in Physics together with the work by
Robert Hofstadter \cite{key-37} on electron scattering in atomic
nuclei. The Mössbauer effect consists in resonant and recoil-free
emission and absorption of gamma rays, without loss of energy, by
atomic nuclei bound in a solid. Previous experiments had shown the
emission and absorption of X-rays. Thus, researchers expected that
an analogous phenomenon could exist also for gamma rays. Differently
from X-rays, which are due to electronic transitions, gamma rays are
due to nuclear transitions. Experiment attempting to observe nuclear
resonance due to gamma-rays in gases failed because of the loss of
energy due to recoil. This indeed prevents resonance. The idea of
Mössbauer was to observe resonance in nuclei of solid iridium \cite{key-36}
by raising the issue of why gamma-ray resonance is possible in solids,
but fails in gases. Mössbauer showed that, if one considers atoms
bound into a solid, with particular assumptions a fraction of the
nuclear events occurs with no recoil \cite{key-36}. Hence, the observed
resonance is due to such a recoil-free fraction of nuclear events
\cite{key-36}. For this reason, the Mössbauer effect is also called
\emph{recoilless nuclear resonance fluorescence}.

Here, we will consider the \emph{Mössbauer rotor effect}, see Figure
1. In that case, one considers an absorber which orbits around the
source of resonant radiation. Assuming that the $z-axis$ is perpendicular
to the plane of Figure 1, the apparatus within the circumference rotates
around such a $z-axis$ with constant angular velocity $\omega$.
The final detector in the right of the Figure is at rest instead.
The experiment permits to measure the so called \emph{transverse Doppler
effect} through the fractional energy shift for a resonant absorber
\cite{key-16}. In fact, through its motion, such a transverse Doppler
effect generates a relative energy shift between emission and absorption
lines. 
\begin{figure}
\includegraphics{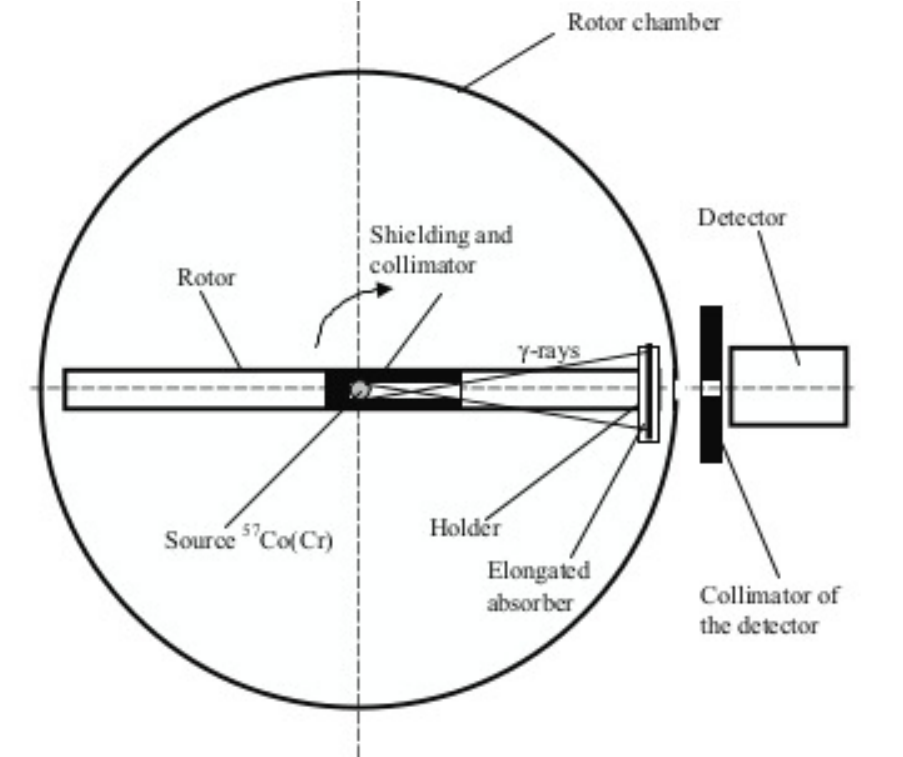}
\begin{description}
\item [{Figure}] \textbf{1} Scheme of the Mössbauer rotor experiment, adapted
from \cite{key-39}. Assuming that the $z-axis$ is perpendicular
to the plane of the Figure, the apparatus within the circumference
rotates around such a $z-axis$ with constant angular velocity $\omega$.
The final detector in the right of the Figure is at rest instead. 
\end{description}
\end{figure}

A historical, important experiment on the Mössbauer rotor effect was
due to Kündig \cite{key-16}. He measured the shift of the 14.4-keV
Mössbauer absorption line of $Fe^{57}$ as a function of the angular
velocity of the rotor \cite{key-16}. He claimed that the transverse
Doppler shift was in agreement with the prevision of the theory of
relativity within an experimental error of $1.1\%,$ by also discussing
possible sources of systematic errors \cite{key-16}. 

A team of researchers recently reanalysed Kündig's experiment \cite{key-38,key-39}.
They started to reanalyse in \cite{key-38} the original data of Kündig's
paper. Then, they decided to realize their own experiment on the transverse
Doppler effect in the Mössbauer rotor framework \cite{key-39}. In
their first work \cite{key-38}, the authors found that the data processing
of Kündig's original analysis was erroneous due to the presence of
mistakes. After the correction of such mistakes, the experimental
data gave the correct value of the fractional energy shift as \cite{key-38}

\begin{equation}
\frac{\bigtriangleup E}{E}=-k\frac{v^{2}}{c^{2}}\label{eq: k}
\end{equation}
Thus, they obtained $k=0.596\pm0.006$ rather than previous result
$k=0.5$ which was considered consistent with the relativity theory
by Kündig \cite{key-16}, Pauli \cite{key-12} and others. This appeared
very strange, but we will see that it can be solved by using both
a special relativistic and a general relativistic treatment. In any
case, from the experimental point of view we stress that the authors
of \cite{key-38} found that
\begin{itemize}
\item the deviation of the coefficient $k$ in the above Eq. (\ref{eq: k})
from $0.5$ is almost 20 times higher than the measuring error;
\item such a deviation is not due to the influence of rotor vibrations and/or
other kinds of disturbing factors. In fact, it seems that all the
potential disturbing factors were excluded by the very good methodology
applied by Kündig in \cite{key-16}. 
\end{itemize}
Kundig's methodology worked through a first-order Doppler modulation
of the energy of $\gamma-$quanta on the rotor at each fixed frequency
of rotation \cite{key-16}. Hence, the experiment of Kündig must be
considered as being very precise and surely more precise than other
similar experiments {[}40\textendash 44{]}. In fact, the experiments
analysed in refs. {[}40\textendash 44{]} measured only the count rate
of detected $\gamma-$quanta as a function of rotation frequency.
The authors of \cite{key-38} have also shown that the experiment
analysed in \cite{key-44} is in agreement with the supposition $k>0.5.$
Such an experiment contains much more data than the others discussed
in {[}40\textendash 43{]}. The new results in \cite{key-38} motivated
the authors to realize their own experiment \cite{key-39}. In this
new experiment, the authors followed neither the scheme of the Kündig's
experiment \cite{key-16}, nor the schemes of the other previously
cited experiments {[}40\textendash 44{]}. In that way, they obtained
a completely independent information on the value of $k$ in Eq. (\ref{eq: k}).
In fact, the authors of \cite{key-39} refrained from the first-order
Doppler modulation of the energy of $\gamma-$quanta. Thus, the uncertainties
in the realization of this method have been excluded \cite{key-39}.
The authors measured the count rate of detected $\gamma-$quanta $N$
as a function of the rotation frequency $\nu$ \cite{key-39}. In
addition, differently from the experiments {[}40\textendash 44{]},
in \cite{key-39} the influence of chaotic vibrations on the measured
value of $k$ has been taken into due account. The authors of \cite{key-39}
also developed a method involving a joint processing of the data collected
for two selected resonant absorbers with the specified difference
of resonant line positions in the Mössbauer spectra. The final result
of the value of $k$ was $k=0.68\pm0.03$ \cite{key-39}. Hence, the
experiment \cite{key-39}. confirmed that the coefficient $k$ in
Eq. (\ref{eq: k}) substantially exceeds $0.5.$ The scheme of the
new Mössbauer rotor experiment is shown in Figure 1,  see \cite{key-39}
for details. 

\section{The Mössbauer rotor effect in the general relativistic framework}

\subsection{The ``gravitational redshift''}

Following \cite{key-13,key-14,key-17}, one considers a transformation
from an inertial coordinate system, centered in the source in Figure
1, with the $z-axis$ perpendicular to the plane of the Figure, to
a second coordinate system, which rotates around the $z-axis$ in
cylindrical coordinates. In the flat Lorentzian coordinate system,
the metric is \cite{key-13,key-14,key-17}

\begin{equation}
ds^{2}=c^{2}dt^{2}-dr^{2}-r^{2}d\phi^{2}-dz^{2}.\label{eq: Minkowskian}
\end{equation}
One performs a transformation to a reference frame $\left\{ t',r',\phi'z'\right\} ,$
which has constant angular velocity $\omega$ around the $z-axis,$
obtaining \cite{key-13,key-14,key-17}

\begin{equation}
\begin{array}{cccc}
t=t'\; & r=r' & \;\phi=\phi'+\omega t'\quad & z=z'\end{array}.\label{eq: trasformazione Langevin}
\end{equation}
Thus, one gets the well known Langevin line-element for the rotating
reference frame \cite{key-7,key-8,key-13,key-14,key-17}
\begin{equation}
ds^{2}=\left(1-\frac{r'^{2}\omega^{2}}{c^{2}}\right)c^{2}dt'^{2}-2\omega r'^{2}d\phi'dt'-dr'^{2}-r'^{2}d\phi'^{2}-dz'^{2}.\label{eq: Langevin metric}
\end{equation}
Through the above discussed EEP, the metric (\ref{eq: Langevin metric})
can be interpreted as representing a stationary gravitational field
\cite{key-13,key-14,key-17}. For a well understanding of technical
details of the last statement the reader can see paragraph 89 of \cite{key-10}
and \cite{key-45}. Then, the EEP permits to consider the inertial
force that a rotating observer experiences as if the same observer
is subjected to a gravitational ``force'' \cite{key-13,key-14,key-17}.
Therefore, a first contribution is due to the ``gravitational blueshift''
\cite{key-13,key-14,key-17}. This contribution can be calculated
starting from Eq. (25.26) in \cite{key-46} as it is written down
in the $20^{th}$ printing of 1997 
\begin{equation}
z\equiv\frac{\Delta\lambda}{\lambda}=\frac{\lambda_{received}-\lambda_{emitted}}{\lambda_{emitted}}=|g_{00}(r'_{1})|^{-\frac{1}{2}}-1.\label{eq: z  MTW}
\end{equation}
Eq. (\ref{eq: z  MTW}) gives the redshift of a photon emitted by
an atom at rest in a gravitational field and received by an observer
at rest at infinity. In the current analysis Eq. (\ref{eq: z  MTW})
must be rescaled. In fact, Eq. (\ref{eq: z  MTW}) applies to a gravitational
field which decreases with increasing radial coordinate, but here
one has a ``gravitational field'' which increases with increasing
radial coordinate $r'$. The result is a \emph{blueshift} (a negative
shift) rather than the typical redshift of a real gravitational field.
In addition, in Eq. (\ref{eq: z  MTW}) the zero potential is set
at infinity, while in the current analysis the zero potential is set
in $r'=0$. One can also use the proper time $\tau$ instead of the
wavelength $\lambda$ by recalling that the delay of the emitted (received)
radiation is connected to the emitted (received) wavelength by $c\Delta\tau=\lambda.$

Therefore, from Eq. (\ref{eq: Langevin metric}), one obtains \cite{key-13,key-14,key-17}
\begin{equation}
\begin{array}{c}
z_{1}\equiv\frac{\Delta\tau_{2}-\Delta\tau_{1}}{\Delta\tau_{1}}=1-|g_{00}(r'_{1})|{}^{-\frac{1}{2}}=1-\frac{1}{\sqrt{1-\frac{\left(r'_{1}\right)^{2}\omega^{2}}{c^{2}}}}\\
\\
=1-\frac{1}{\sqrt{1-\frac{v^{2}}{c^{2}}}}\simeq-\frac{1}{2}\frac{v^{2}}{c^{2}}.
\end{array}\label{eq: gravitational redshift}
\end{equation}
Strictly speaking, Eqs. (\ref{eq: z  MTW}) and (\ref{eq: gravitational redshift})
are valid only for time-independent metrics with $g_{0j}=0$ \cite{key-46}.
In general, this is not the case of Eq. (\ref{eq: Langevin metric}).
On the other hand, resonant radiation propagates from the source in
the radial direction in the laboratory frame. For radial motions it
is $d\phi=0$ in the laboratory frame. Therefore, one gets from Eq.
(\ref{eq: trasformazione Langevin}) 
\begin{equation}
d\phi'+\omega dt'=0\Rightarrow\omega=-\frac{d\phi'}{dt'}.\label{eq: light propagation}
\end{equation}
Then, for radial motions in the laboratory frame, the Langevin metric
of Eq. (\ref{eq: Langevin metric}) reduces to 
\begin{equation}
ds^{2}=\left(1-\frac{r'^{2}\omega^{2}}{c^{2}}\right)c^{2}dt'^{2}-dr'^{2}+r'^{2}d\phi'^{2}-dz,\label{eq: Langevin corretta}
\end{equation}
which has indeed $g_{0i}=0$ and enables the using of Eqs. (\ref{eq: z  MTW})
and (\ref{eq: gravitational redshift}) in the present approach. In
Eq. (\ref{eq: gravitational redshift}) $\Delta\tau_{1}$ is the delay
of the emitted radiation, $\Delta\tau_{2}$ is the delay of the received
radiation, $r'_{1}=c\tau_{1}$ is the radial coordinate of the detector,
where $\tau_{1}$ is the proper time that has been measured by the
rotating observer during the trajectory of the light, and $v=r'_{1}\omega$
is the tangential velocity of the detector \cite{key-13,key-14,key-17}.
The physical meaning of $\Delta\tau_{1}\neq\Delta\tau_{2}$ is that
the proper time between two subsequent photons emissions as recorded
by the clock at the source is different from the proper time between
the subsequent absorptions of the same two photons as recorded by
the identical clock at the absorber. This was the original way of
thinking of Einstein \cite{key-47}, adopted in an elegant way by
Weyl \cite{key-48}. In other words, $\Delta\tau_{1}$ denotes the
proper time at the source that elapses between successive light-ray
emissions (or, say, the emission of a single wavelength), and $\Delta\tau_{2}$
denotes the proper time elapsing on the absorber between successive
light-ray absorptions. On the other hand, the above analysis has been
made in rotating frame. But the final measurement is made in the Lorentzian
frame in the laboratory. In such a frame the static observer sees
no ``gravitational field''. Instead, the Lorentzian observer sees
the opposite \emph{redshift} predicted by special relativity when
the final detector sees the absorber as being at its closest approach.
Kündig \cite{key-16} pointed out that the rotating observer comes
to the conclusion that his clock is slowed down by the ``gravitational
potential''. Thus, the clock of the Lorentzian observer is faster
than the clock of the rotating observer. Then, the fractional energy
shift in the laboratory results 
\begin{equation}
\frac{E_{2}-E_{1}}{E_{1}}=\frac{\triangle E_{1}}{E_{1}}=\simeq-\frac{1}{2}\frac{v^{2}}{c^{2}}.\label{eq: fractional energy shift}
\end{equation}
Hence, one gets $k_{1}=\frac{1}{2}$ as the contribution to $k$ from
the first effect \cite{key-13,key-14,key-17}. Following Pauli's book
\cite{key-12}, Kündig \cite{key-16} obtained the same result of
$k_{1}=\frac{1}{2}$. 

\subsection{The ``cosmological redshift''}

The necessity of an additional effect can be understood through the
following observation. The calculation in Subsection 3.1 has been
made in the reference frame of the rotating observer. As the final
detector moves with respect to the rotating observer \cite{key-13,key-14,key-17},
see Figure 1, the clock of the Lorentzian non-rotating observer is
\emph{not} synchronized with the clock of the rotating observer. Therefore,
a second, additional effect contributes to the transverse Doppler
effect \cite{key-13,key-14,key-17}. This is an effect of clock synchronization
which was not present in previous works \cite{key-16,key-38,key-39}.
This second effect was erroneously calculated in \cite{key-13,key-14}
by using Eq. (10) of \cite{key-32}. Such an equation represents the
variation of proper time $d\tau$ on the moving clock having radial
coordinate $r'$ for values $v\ll c$. We wrote ``erroneously''
for the following reasons. Eq. (10) of \cite{key-32} reads

\begin{equation}
d\tau=dt'-\frac{\omega r'^{2}d\phi'}{c^{2}}.\label{eq:secondo contributo}
\end{equation}
Instead, in \cite{key-13,key-14} it has been used 
\begin{equation}
d\tau=dt'\left(1-\frac{r'^{2}\omega^{2}}{c^{2}}\right).\label{eq:secondo contributo sbagliata}
\end{equation}
Thus, one sees that Eqs. (\ref{eq:secondo contributo}) and (\ref{eq:secondo contributo sbagliata})
are equal only if $\omega=\frac{d\phi'}{dt'}.$ But this is not the
case because in \cite{key-13,key-14} light which propagates in the
radial direction has been considered. This should imply $d\phi'=dz'=0$,
but this is another mistake because in the rotating frame light \emph{does
not} propagate in the radial direction. Let us perform a correct analysis.
Following \cite{key-17}, one observes that $t'$ represents the time
coordinate for the rotating observer. But Eq. (\ref{eq: trasformazione Langevin})
means also $t'=t$, and $t$ represents both of the coordinate time
and the proper time for the inertial observer located in the laboratory.
The interpretation of the rotating frame in terms of a gravitational
field permits to relate the rate $d\tau$ of the proper time to the
rate $dt'$ of the coordinate time as \cite{key-10,key-17}
\begin{equation}
d\tau^{2}=g_{00}dt'^{2}.\label{eq: relazione temporale}
\end{equation}
Eq. (\ref{eq: relazione temporale}) works for any two infinitesimally
separated events occurring at one and same point in space. Now, Eq.
(\ref{eq: Langevin metric}) gives $g_{00}=\left(1-\frac{r'^{2}\omega^{2}}{c^{2}}\right).$
Therefore, Eq. (\ref{eq: relazione temporale}) becomes 
\begin{equation}
c^{2}d\tau^{2}=\left(1-\frac{r'^{2}\omega^{2}}{c^{2}}\right)c^{2}dt'^{2}.\label{eq: relazione temporale 2}
\end{equation}
Hence, by using again Eq. (\ref{eq: trasformazione Langevin}), one
obtains
\begin{equation}
c^{2}dt'^{2}=c^{2}dt{}^{2}=dr^{2}=dr'^{2},\label{eq: obviously}
\end{equation}
where the equality 
\begin{equation}
c^{2}dt{}^{2}=dr^{2}\label{eq: equality}
\end{equation}
is obtained because light propagates in the radial direction for the
observer in the laboratory (the source is indeed at rest in the laboratory
frame). Thus, one can set $d\phi=dz=0$ in Eq. (\ref{eq: Minkowskian})
and, by inserting the condition of null geodesics $ds=0$ in the same
equation one gets Eq. (\ref{eq: equality}). Therefore, Eq. (\ref{eq: relazione temporale 2})
becomes 
\begin{equation}
c^{2}d\tau^{2}=\left(1-\frac{r'^{2}\omega^{2}}{c^{2}}\right)dr'^{2}.\label{eq: relazione temporale 3}
\end{equation}
Taking the root square of Eq. (\ref{eq: relazione temporale 3}) one
gets 

\begin{equation}
cd\tau=\sqrt{1-\frac{r'^{2}\omega^{2}}{c^{2}}}dr'.\label{eq: secondo contributo finale}
\end{equation}
Eq. (\ref{eq: secondo contributo finale}) coincides with Eq. (10)
in \cite{key-13}. Hence, now one can carefully remake the analysis
in \cite{key-13,key-14}. One can approximate Eq. (\ref{eq: secondo contributo finale})
with
\begin{equation}
cd\tau\simeq\left(1-\frac{1}{2}\frac{r'^{2}\omega^{2}}{c^{2}}+....\right)dr'.\label{eq: well approximated}
\end{equation}
Eq. (\ref{eq: well approximated}) takes into account the second effect
of order $\frac{v^{2}}{c^{2}}$ to (proper) time dilation
\begin{equation}
c\Delta\tau_{3}=\int_{0}^{r'_{1}}\left(1-\frac{1}{2}\frac{\left(r'\right)^{2}\omega^{2}}{c^{2}}\right)dr'-r'_{1}=-\frac{1}{6}\frac{\left(r'_{1}\right)^{3}\omega^{2}}{c^{2}}=-\frac{1}{6}r'_{1}\frac{v^{2}}{c^{2}}.\label{eq: delta tau 2}
\end{equation}
$\Delta\tau_{3}$ in this equation represents the difference between
the proper time (distance) that has been measured by the rotating
observer and the proper time (distance) that has been measured by
the Lorentzian observer during the trajectory of the light which propagates
from the source to the final detector. Then, the additional effect
of clock synchronization (at order $\frac{v^{2}}{c^{2}}$) to the
blueshift is 
\begin{equation}
z_{2}\equiv\frac{\Delta\tau_{3}}{\tau_{1}}=-k_{2}\frac{v}{c^{2}}^{2}=-\frac{1}{6}\frac{v^{2}}{c^{2}}.\label{eq: z2}
\end{equation}
This means $k_{2}=\frac{1}{6}$. This effect is something similar
to the cosmological redshift. In fact, Eq. (\ref{eq: delta tau 2})
shows that a variation of the \emph{proper distance (time)} between
the source and the detector is present, while the radial distance,
in the sense of the difference of the radial coordinates of the source
and the detector, remains constant. Thus, in a certain sense, the
radial coordinate $r'$ represents a \emph{comoving} \emph{coordinate}.
In fact, if one defines the proper time that has been measured by
the Lorentzian observer during the trajectory as $\tau_{L}\equiv\tau_{1}$
and the proper time that has been measured by the rotating observer
during the trajectory as $\tau_{R}\equiv\Delta\tau_{3}+\tau_{Lor},$
Eq. (\ref{eq: delta tau 2}) can be rewritten as 
\begin{equation}
1+z_{2}=\frac{\tau_{R}}{\tau_{L}},\label{eq: z2+1}
\end{equation}
which is similar to the well known formula of the cosmological redshift
\cite{key-46}
\[
1+z=\frac{a_{r}}{a_{e}},
\]
where $a_{r}$ and $a_{e}$ are the values of the scale factor of
the Universe at the instants of the reception and of the emission
of the light, respectively. This has an intuitive and intriguing representation
if one recalls that in the rotating coordinate system the ratio of
the circumference to its radius in a plane with $z=constant$ is larger
than $2\pi$ \cite{key-10}. In fact, the element of spacial distance
for the Langevin metric (\ref{eq: Langevin metric}) is \cite{key-10}
\begin{equation}
dl^{2}=dr'^{2}+dz'^{2}+\frac{r'^{2}d\phi'^{2}}{1-\frac{r'^{2}\omega^{2}}{c^{2}}},\label{eq: spacial distance}
\end{equation}
which immediately shows that the ratio of the circumference to its
radius in a plane with $z=constant$ is $\frac{2\pi}{\sqrt{1-\frac{r'^{2}\omega^{2}}{c^{2}}}}$.
Thus, in a certain sense the rotating observer ``lives'' in a space
which is contracted in its proper distance (time) in the radial direction
perpendicular to the axis of rotation. This is the physical meaning
of Eq. (\ref{eq: delta tau 2}). 

The following analysis shows that Eqs. (\ref{eq: z2}) and (\ref{eq: z2+1})
really represent a blueshift. Let us consider the first peak of wave
which arrives at the detector after the activation of the source of
light. Then, the Lorentzian observer will count a number of wavelengths
\begin{equation}
N_{L}\equiv\frac{c\tau_{1}}{\lambda_{L}},\label{eq: lunghezze d'onda L}
\end{equation}
where $\lambda_{L}$ is the wavelength of light. Instead, the rotating
observer will count a number of wavelengths 
\begin{equation}
N_{R}\equiv\frac{c\tau_{1}+c\Delta\tau_{3}}{\lambda_{R}},\label{eq: lunghezze d'onda R}
\end{equation}
where now the wavelength of light is $\lambda_{R}$. Causality requires
$N_{R}=N_{R}$ which implies 
\begin{equation}
\frac{\lambda_{R}}{\lambda_{L}}=\frac{c\tau_{1}+\Delta\tau_{3}}{c\tau_{1}}=\frac{\tau_{R}}{\tau_{L}},\label{eq: lunghezze d'onda finale}
\end{equation}
which is equal to Eq. (\ref{eq: z2+1}). 

Also in this case the analysis has been made in the rotating frame.
But, again, the final measurement is made in the Lorentzian frame
in the laboratory. Thus, the observer in the Lorentzian frame will
measure an opposite effect of clock synchronization $-c\Delta\tau_{3}$
resulting in a redshift expressed by the inverse wavelengths ratio
\begin{equation}
\frac{\lambda_{L}}{\lambda_{R}}=\frac{\tau_{L}}{\tau_{R}},\label{eq: redshift}
\end{equation}
which corresponds to a second fractional energy shift 
\begin{equation}
\frac{\triangle E_{2}}{E_{1}}\simeq-\frac{1}{6}\frac{v^{2}}{c^{2}}.\label{eq: fractional energy shift2}
\end{equation}
Therefore, Eqs. (\ref{eq: fractional energy shift}) and (\ref{eq: fractional energy shift2})
permit to obtain the total fractional energy shift as
\begin{equation}
\frac{\triangle E}{E_{1}}=\frac{\triangle E_{1}}{E_{1}}+\frac{\triangle E_{2}}{E_{1}}\simeq-\frac{2}{3}\frac{v^{2}}{c^{2}}.\label{eq: z totale}
\end{equation}
This theoretical result is completely consistent with the experimental
result $k=0.68\pm0.03$ in \cite{key-39}. 

\section{Discussion}

Let us discuss the physical meaning of the above general relativistic
analysis. The additional factor $-\frac{1}{6}$ in Eq. (\ref{eq: fractional energy shift2})
is due to clock synchronization. This generates a further redshift
effect that in a certain sense is analogous to the cosmological redshift,
because it is due to the variation of the proper distance (time) between
the source and the detector in the rotating frame. Missing its presence
in \cite{key-16,key-35,key-38,key-39,key-49,key-50} was due to the
incorrect comparison of proper time rates between the clock of the
rotating observer and the clock of the Lorentzian observer. Consequences
of this misunderstanding were claims of invalidity of the GTR and/or
some unscientific attempt to explain the Mössbauer rotor through ``exotic''
effects \cite{key-49,key-50}. Such unscientific effects must be instead
completely rejected. 

Reference \cite{key-32} is useful for analysing the Langevin line-element.
This is similar to the use of the GTR in Global Positioning Systems
(GPS) and permits to gain an interesting physical meaning \cite{key-13,key-14}:
the additional effect giving the correction of $-\frac{1}{6}$ in
Eq. (\ref{eq: z2}) is analogous to the additional effect that one
must consider in GPS when one takes into account the difference between
the time measured in a frame co-rotating with the Earth geoid and
the time measured in a non-rotating (locally inertial) Earth centered
frame (and also the difference between the proper time of an observer
at the surface of the Earth and at infinity). In fact, one cannot
merely consider the gravitational redshift due to the Earth's gravitational
field, without considering the effect of the Earth's rotation. In
that case, GPS cannot work \cite{key-13,key-14}! The insight is that
one cannot merely use the time elapsing on the orbiting GPS clocks
in order to transfer time from one transmission event to another.
Instead, one must be careful to consider important path-dependent
effects. This is exactly what happened in the above analysis of clock
synchronization \cite{key-13,key-14}. In other words, the additional
effect giving the important correction $-\frac{1}{6}$ in Eq. (\ref{eq: z2})
must be considered neither an obscure mathematical artifact nor a
negligible physical detail. Instead, it is a remarkable physical effect
that one must take into due account \cite{key-13,key-14}. Further
details on the analogy between the above analysis and the use of the
GTR in GPS have been highlighted in \cite{key-13,key-14}.

For the sake of completeness, we stress that our results are also
confirmed by two recent results on the Mössbauer rotor experiment
\cite{key-15,key-51}. In \cite{key-15} the authors wanted to underline
some mathematical aspects in general relativistic frameworks. They
neither discuss in detail the different physical interpretations of
the experimental results proposed during the recent years nor wanted
to propose a new one. Instead, starting from the analysis in \cite{key-13},
they analyzed three different types of time involved in the Mössbauer
rotor experiment, linking a term introduced in \cite{key-13} to the
difference between coordinate and physical velocity of light. In \cite{key-51}
the approach has been considered from the viewpoint of arbitrary moving
continuum. Continuum characteristics have been determined by a deformation
tensor, a stress tensor, a strain velocity tensor, a spin tensor and
the first curvature vectors of the world lines of medium particles
\cite{key-51}. It has been indeed shown that these physical values
are located on the hypersurfaces orthogonal to the world lines of
medium particles and specify the physical space \cite{key-51}. The
time counting from the physical space permits to obtain the blueshift
of the frequency at the end of the tube that coincides with the above
result \cite{key-51}.

Another recent result in \textit{\emph{\cite{key-18}}}\textit{ }\textit{\emph{has
shown that a general relativistic analysis similar to the one proposed
in this paper works also for the Sagnac effect. }}

\section{Clock synchronization is a real effect}

In the recent paper \cite{key-35}, the Authors claim that the second
effect of clock synchronization does not exist. The Authors of \cite{key-35}
criticize indeed the results in the paper of Iovane and Benedetto,
i.e. Ref. \cite{key-15}. In fact, such results are consistent with
our previous results in \cite{key-13,key-14,key-17} and with the
analysis in Section 3.2 of this paper, because in \cite{key-15} it
is shown that the Mössbauer rotor apparatus substantially generates
a desynchronization of clocks between the reference frame of laboratory
and the reference frame of the rotating observer. Consequently, the
two reference frames must be re-synchronized as Section 3.2 of this
paper and in \cite{key-13,key-14,key-17}. The criticism of the Authors
of \cite{key-35} is that, in their paper, verbatim, 

\emph{following Corda, Iovane and Benedetto} \emph{considered the
propagation of photons in the radial direction of the rotating system,
and further defined the proper time of the absorber next to the proper
time of the detector, and they finally derived the expected energy
shift between emission and absorption lines in Mössbauer rotor experiments
involving the so-called effect of the ``desynchronization of clocks''
that they have introduced. }

Then, the Authors of \cite{key-35} add that, verbatim, 

\emph{However, the fact remains that in all known Mössbauer experiments
in a rotating system implemented up to date (including {[}15, 18-21,
23-27{]}), the resonant -quanta propagate over the rotor along the
straight line, joining a spinning source and a detector as seen by
a laboratory observer. }

Consequently, the Authors of \cite{key-35} conclude that 

\emph{the problem of the correct interpretation of the results of
Mössbauer experiments in a rotating system cannot be reduced to an
unaccounted-for desynchronization effects between the clocks in a
rotating system and in a laboratory frame - so much so that further
search of possible approaches to the explanation of these experiments
in the framework of general theory of relativity (GTR) is required.}

Here we do not want to enter in the details of the paper of Iovane
and Benedetto \cite{key-15} but we must stress that, concerning our
analysis, the Authors of \cite{key-35} make confusion. In fact, one
can easily check in Section 3.2 of this paper as well as in \cite{key-17}
that we considered light propagation \textbf{in the radial direction
of the observer in the laboratory, }not in the radial direction of
the rotating observer. Thus, the criticism of the Authors of \cite{key-35}
is completely inconsistent. 

On the other hand, the claim in \cite{key-35} can be easily dismissed
with the following analysis. Let us rewrite Eq. (\ref{eq: relazione temporale})
as
\begin{equation}
d\tau^{2}=g_{00}dt{}^{2}.\label{eq: relazione temporale finale}
\end{equation}
We recall that this equation works for any two infinitesimally separated
events occurring at one and the same point in space. In the Lorentzian
frame one has $g_{00}=1.$ Then 
\begin{equation}
d\tau_{L}^{2}=dt_{L}^{2},\label{eq: relazione temporale finale laboratorio}
\end{equation}
where we re-labelled with $\tau_{L}$ and $t_{L}$ the proper time
and the coordinate time of the observer in the Lorentzian frame. In
the case of the rotating clock one has $g_{00}=\left(1-\frac{r'^{2}\omega^{2}}{c^{2}}\right).$
Then 
\begin{equation}
d\tau_{R}^{2}=\left(1-\frac{r'^{2}\omega^{2}}{c^{2}}\right)dt_{R}^{2},\label{eq: relazione temporale finale rotazione}
\end{equation}
where we re-labelled with $\tau_{R}$ and $t_{R}$ the proper time
and the coordinate time of the rotating observer. But, from the Langevin
transformation (\ref{eq: trasformazione Langevin}) it is 
\begin{equation}
t_{R}=t_{L}.\label{eq: uguaglianza}
\end{equation}
 Then, from the last 3 equations one immediately gets 
\begin{equation}
d\tau_{R}^{2}=\left(1-\frac{r'^{2}\omega^{2}}{c^{2}}\right)d\tau_{L}^{2},\label{eq: relazione temporale finalissima}
\end{equation}
which means that clock of the rotating observer and the clock of the
Lorentzian observer in the laboratory measure the same proper time
\textbf{if and only if} $r=r'=0,$ i.e. only in the origin of the
two reference frames. Thus, clock synchronization is necessary in
all the other points. Hence, Eq. (\ref{eq: relazione temporale finalissima})
gives the difference of proper time between the rotating observer
and the observer at rest in the laboratory frame. Thus, we have shown
that the claim in \cite{key-35} that the second effect of clock synchronization
does not exist is wrong.

\section{Conclusion remarks}

Both the history of the Mössbauer rotor effect and its interpretation
from the point of view of Einstein's GTR have been analysed by adding
some new insight. In particular, it has been shown that, if on one
hand the ``traditional'' effect of redshift has a strong analogy
with the gravitational redshift, the additional effect of redshift
due to clock synchronization has an intriguing analogy with the cosmological
redshift. The general relativistic interpretation of the Mössbauer
rotor experiment is an important consequence of the EEP. In Section
5 of this paper it has been shown that the claim in \cite{key-35}
that the second effect of clock synchronization does not exist is
not correct. 

\section{Acknowledgements}

The Author thanks an unknown Referee for very useful comments and
suggestions.


\begin{thebibliography}{10}
\bibitem[1]{key-1}G. Sagnac, \emph{L'ether lumineux demontre par
l'effet du vent relatif d'ether dans un interferometre en rotation
uniforme}. Comptes Rendus, \textbf{157}, 708 (1913). 

\bibitem[2]{key-2}G. Sagnac, \emph{Sur la preuve de la realite de
l'ether lumineux par l'experience de l'interferographe tournant. Par
l'effet du vent relatif d'ether dans un interferometre en rotation
uniforme}. Comptes Rendus \textbf{157}, 1410 (1913). 

\bibitem[3]{key-3}M. von Laue, \emph{Zur Diskussion iiber den starren
Korper in der Relativitatstheorie}. Phys. Z \textbf{12}, 85 (1911). 

\bibitem[4]{key-4}A. Einstein, \emph{Bemerkung zu P. Harzers Abhandlung:
Die Mitfuhrung des Lichtes in Glas und die Aberration}. Astr. Nach.
\textbf{199}, 8 (1914).

\bibitem[5]{key-5}A. Einstein, \emph{Antwort auf eine Replik P. Harzers}.
Astr. Nach. \textbf{199}, 47 (1914).

\bibitem[6]{key-6}M. von Laue, \emph{Theoretisches liber optische
Beobachtungen zur Relativitatstheorie}. Phys. Z \textbf{21}, 659 (1920).

\bibitem[7]{key-7}P. Langevin, \emph{Sur la theorie de relativite
et l'experience de M. Sagnac}. Comptes Rendus, \textbf{173}, 831 (1921).

\bibitem[8]{key-8}P. Langevin, \emph{Relativite - Sur l'experience
de Sagnac}. Comptes Rendus \textbf{2015}, 304 (1937).

\bibitem[9]{key-9}G. Rizzi and M. L. Ruggiero, \emph{The relativistic
Sagnac Effect: two derivations}. In G. Rizzi; M.L. Ruggiero (eds.).
Relativity in Rotating Frames. Dordrecht: Kluwer Academic Publishers
(2003).

\bibitem[10]{key-10}L. D. Landau and E. M. Lifshitz, \emph{The Classical
Theory of Fields}. 2nd edition, Pergamon Press, (1962).

\bibitem[11]{key-11}A. Einstein, \emph{The Collected Papers}, Volume
6: The Berlin Years: Writings, 1914-1917 (English translation supplement)
Pages 31-32. 

\bibitem[12]{key-12}W. Pauli, \emph{Theory of Relativity, }Pergamon
Press, London (1958).

\bibitem[13]{key-13}C. Corda, \emph{New proof of general relativity
through the correct physical interpretation of the Mossbauer rotor
experiment}. Int. Journ. Mod. Phys. D \textbf{27}, 1847016 (2018). 

\bibitem[14]{key-14}C. Corda, \emph{Interpretation of Mossbauer experiment
in a rotating system: a new proof for general relativity}. Ann. Phys.
\textbf{355}, 360 (2015). 

\bibitem[15]{key-15}G. Iovane, E. Benedetto, \emph{Coordinate velocity
and desynchronization of clocks}. Ann. Phys. \textbf{403}, 106 (2019). 

\bibitem[16]{key-16}\emph{W. Kündig, Measurement of the Transverse
Doppler Effect in an Accelerated System}. Phys. Rev. \textbf{129},
2371 (1963).

\bibitem[17]{key-17}C. Corda, \emph{Mossbauer rotor experiment as
new proof of general relativity: Rigorous computation of the additional
effect of clock synchronization}. Int. Journ. Mod. Phys. D \textbf{28},
1950131 (2019).

\bibitem[18]{key-18}E. Benedetto, F. Feleppa, I. Licata, H. Moradpour,
C. Corda, \emph{On the general relativistic framework of the Sagnac
effect}. Eur. Phys. J. C \textbf{79}, 187 (2019).

\bibitem[19]{key-19}H. Lichtenegger and B. Mashhoon, in second chapter
of \emph{The Measurement of Gravitomagnetism: A Challenging Enterprise},
edited by L. Iorio (Nova Science, New York, 2007). 

\bibitem[20]{key-20-2}D. R. Brill and J. M. Cohen, \emph{Rotating
Masses and Their Effect on Inertial Frames}. Phys. Rev. \textbf{143},
1011 (1966). 

\bibitem[21]{key-21}E. L. Schucking, E. J. Surowitz, \emph{Einstein's
Apple: His First Principle of Equivalence}. arXiv:gr-qc/0703149.

\bibitem[22]{key-22}R. P. Kerr, \emph{Gravitational Field of a Spinning
Mass as an Example of Algebraically Special Metrics}. Phys. Rev. Lett.
\textbf{11}, 237 (1963). 

\bibitem[23]{key-23}E. T. Newman, E. Couch, K. Chinnapared, A. Exton,
A. Prakash, and R. Torrence, \emph{Metric of a Rotating, Charged Mass}.
Jour. Math. Phys. \textbf{6}, 918 (1965). 

\bibitem[24]{key-24} H. Pfister, \emph{On the history of the so-called
Lense-Thirring effect}. Gen. Rel. Grav. \textbf{39}, 1735 (2007). 

\bibitem[25]{key-25}L. Iorio, H. I. M. Lichtenegger, M. L. Ruggiero
and C. Corda, \emph{Phenomenology of the Lense-Thirring effect in
the solar system}. Astrophys. Space Sci. \textbf{331}, 351 (2013). 

\bibitem[26]{key-26}L. Iorio, \emph{The post-Newtonian gravitomagnetic
spin-octupole moment of an oblate rotating body and its effects on
an orbiting test particle; are they measurable in the Solar System?}
Mon. Not. Roy. Astron. Soc. \textbf{484}, 4811 (2019).

\bibitem[27]{key-27}L. Iorio, \emph{Measuring general relativistic
dragging effects in the Earth's gravitational field with ELXIS: a
proposal}. Class. Quant. Gravit. \textbf{36}, 035002 (2019). 

\bibitem[28]{key-28}L. Iorio and C. Corda, \emph{Gravitomagnetism
and Gravitational Waves}. Op. Astr. Jour. Suppl. \textbf{1-M5}, 84
(2011).

\bibitem[29]{key-29}M. L. Ruggiero, A. Tartaglia, \emph{Test of gravitomagnetism
with satellites around the Earth}. Eur. Phys. J. Plus \textbf{134},
205 (2019). 

\bibitem[30]{key-30}M. L. Ruggiero, \emph{Gravitomagnetic Field of
Rotating Rings}. Astrophys. Space Sci. \textbf{361}, 140 (2016). 

\bibitem[31]{key-31}M. L. Ruggiero, \emph{Gravito-electromagnetic
effects of massive rings}. Int. Journ. Mod. Phys. D \textbf{24}, 1550060
(2015). 

\bibitem[32]{key-32}N. Ashby, \emph{Relativity in the Global Positioning
System}. Liv. Rev. Rel. 6, 1 (2003). 

\bibitem[33]{key-33}J. H. Oort, \emph{The force exerted by the stellar
system in the direction perpendicular to the galactic plane and some
related problems}. Bull. Astr. Neth. 6, 249 (1932). 

\bibitem[34]{key-34}F. Zwicky, \emph{Die Rotverschieb ung von extragalaktischen
Nebeln}. Helv. Phys. Acta 6, 110 (1933). 

\bibitem[35]{key-35}A. L. Kholmetskii, T. Yarman, O. Yarman, M. Arik,
\emph{Concerning Mossbauer experiments in a rotating system and their
physical interpretation}. Ann. Phys. \textbf{411}, 167912 (2019). 

\bibitem[36]{key-36}R. L. Mössbauer, \emph{Kernresonanzfluoreszenz
von Gammastrahlung in Ir$^{191}$.} Zeitschrift für Physik A (in German)
\textbf{151}, 124 (1958). 

\bibitem[37]{key-37}R. Hofstadter, \emph{The Electron-Scattering
Method and Its Application to the Structure of Nuclei and Nucleons.
}Nobel Lecture, December 11, 1961.

\bibitem[38]{key-38}A. L. Kholmetskii, T. Yarman and O. V. Missevitch,
\emph{Kündig's experiment on the transverse Doppler shift re-analyzed}.
Phys. Scr. \textbf{77}, 035302 (2008). 

\bibitem[39]{key-39}A. L. Kholmetskii, T. Yarman, O.V. Missevitch
and B. I. Rogozev, \emph{A Mössbauer experiment in a rotating system
on the second-order Doppler shift: confirmation of the corrected result
by Kündig}. Phys. Scr. \textbf{79}, 065007 (2009).

\bibitem[40]{key-40}H. J. Hay et al, \emph{Measurement of the Red
Shift in an Accelerated System Using the Mössbauer Effect in} $Fe^{57}$
Phys. Rev. Lett. \textbf{4}, 165 (1960).

\bibitem[41]{key-41}H. J. Hay, in \emph{Proc. 2nd Conf. Mössbauer
Effect}, ed A Schoen and D M T Compton (New York: Wiley) p 225 (1962). 

\bibitem[42]{key-42}T. E. Granshaw and H. J. Hay, in \emph{Proc.
Int. School of Physics, \textquoteleft Enrico Fermi}\textquoteright{}
(New York: Academic) p 220 (1963). 

\bibitem[43]{key-43}D. C. Champeney and P. B. Moon, \emph{Absence
of Doppler Shift for Gamma Ray Source and Detector on Same Circular
Orbit}. Proc. Phys. Soc. \textbf{77}, 350 (1961).

\bibitem[44]{key-44} D. C. Champeney, G. R. Isaak and A. M. Khan,
\emph{A time dilatation experiment based on the Mössbauer effect}.
Proc. Phys. Soc. \textbf{85}, 583 (1965).

\bibitem[45]{key-45}G. Rizzi and M. L. Ruggiero, \emph{Space Geometry
of Rotating Platforms: An Operational Approach.} Found. Phys. \textbf{32},
1525 (2002). 

\bibitem[46]{key-46} C. W. Misner , K. S. Thorne, J. A. Wheeler,
\emph{\textquotedblleft Gravitation\textquotedblright }, Feeman and
Company (1973). 

\bibitem[47]{key-47}A. Einstein, \emph{Die Grundlage der allgemeinen
Relativitatstheorie}. Ann. Phys. \textbf{354}, 769 (1916). 

\bibitem[48]{key-48}H. Weyl, \emph{Ressource Vorlesungen uber Allgemeine
Relativitatstheorie}. Raum, Zeit, Materie (Verlag von Julius Springer,
Berlin, 1923). 

\bibitem[49]{key-49}T. Yarman, A. L. Kholmetskii, and M. Arik, \emph{Mössbauer
experiments in a rotating system: Recent errors and novel interpretation}.
Eur. Phys. Jour. Plus \textbf{130}, 191 (2015). 

\bibitem[50]{key-50}A. L. Kholmetskii, T. Yarman, O. Yarman, M. Arik,
\emph{Response to ``The Mössbauer rotor experiment and the general
theory of relativity'' by C. Corda.} Ann. Phys. \textbf{374}, 247
(2016). 

\bibitem[51]{key-51}J. Foukzon, E. R. Men'kova, Comment on \emph{``The
Mössbauer rotor experiment and the general theory of relativity''}
\emph{{[}Ann. Physics 368 (2016) 258-266{]}.} Ann. Phys. \textbf{413},
168047 (2020). 
\end{thebibliography}
\end{document}